\documentclass[authoryear,11pt]{elsarticle}

\usepackage{amssymb,amsthm,graphicx,float}

\begin{document}

\begin{frontmatter}

\title{Dimension Reduction of High-Dimensional Datasets Based on Stepwise SVM}

\author{Elizabeth P. Chou, Tzu-Wei Ko}

\address{Department of Statistics, National Chengchi University, No.64, Sec. 2, Zhinan Rd., Wenshan Dist., Taipei City 116, Taiwan}

\begin{abstract}
\textbf{Background}\\

The current study proposes a dimension reduction method, stepwise support vector machine (SVM), to reduce the dimensions of large p small n datasets. The proposed method is compared with other dimension reduction methods, namely, the Pearson product difference correlation coefficient (PCCs), recursive feature elimination based on random forest (RF-RFE), and principal component analysis (PCA), by using five gene expression datasets. Additionally, the prediction performance of the variables selected by our method is evaluated. \\

\noindent\textbf{Results}\\

The study found that stepwise SVM can effectively select the important variables and achieve good prediction performance. Moreover, the predictions of stepwise SVM for reduced datasets was better than those for the unreduced datasets. The performance of stepwise SVM was more stable than that of PCA and RF-RFE, but the performance difference with respect to PCCs was minimal.\\

\noindent\textbf{Conclusions}\\

It is necessary to reduce the dimensions of large p small n datasets. We believe that stepwise SVM can effectively eliminate noise in data and improve the prediction accuracy in any large p small n dataset.
\end{abstract}


\end{frontmatter}


\section{Introduction}
\label{sec1}
Along with the advancements in technology, datasets are produced in more complex structures that classic data clustering methods cannot handle  \citep{de2013analysis}. Therefore, researchers are facing multiple challenges, especially in high-dimensional data learning. High-dimensional covariate information provides a detailed description of individuals involved in a machine learning and classification problem. It provides more information and opportunities from a learning perspective but results in computational challenges as well. One of the major problems with high-dimensional datasets is that it is difficult to determine whether all the variables are important to understanding the underlying phenomena of interest \citep{Fodor02DRSurvey}.If too many covariates are used, the model may be overestimated and either fail to converge or yield an unstable result. These issues are particularly pressing in a large $p$ small $n$ data framework, which is common in genetics and computational biology, for example, microarray analysis. How to identify a target subset from thousands of gene expressions from thousands to diagnose a disease of interest is an increasingly popular and important question. In resolving high-dimensional classification issues, substantial research effort and attention have been directed to dimension reduction to enable the application of classic supervised learning rules \citep{Pu2016, Cai2017, Gottlieb2015, 7782082}. Dimension reduction can improve the quality of the classification and clustering techniques in machine learning. Various fields use different names for the covariate dimensions. In the statistics literature, researchers call covariates ``variables", and in computer science and machine learning, researcher call them ``features" and ``attributes". The goal of dimension reduction or feature selection is to obtain a subset of variables from the original dataset that maintain the majority of the original geometric information  \citep{Fodor02DRSurvey}.

Dimension reduction can be approached in two ways: feature selection or feature extraction. Feature selection reduces the irrelevant dimensions to obtain better classification results with the reduced data. Feature selection methods can be categorized into three types, filter, wrapper, and embedded, based on how the feature selected method interacts with a classifier. In filter methods, variables are ranked according to relevancy, and a threshold is created as a criterion in variable selection. Some commonly used scoring methods are correlation and mutual information. Wrapper methods are classifier dependent. Suboptimal subsets of variables are found by evaluating the predictor performance. Some sequential selection techniques and heuristic search algorithms (e.g., genetic algorithm) are wrapper methods. Embedded methods are also classifier dependent; however, embedded methods consume less computation time than wrapper methods because they perform classification with feature selection simultaneously and select an optimal subset of predictors. These methods, such as decision tree and Lasso, are popular and easy to apply with the built-in algorithm packages in many statistical programming software \citep{Blum1997, Guyon2003,  Saeys2007, Chandrashekar2014}. Through feature extraction, researchers alter the original values of variables to reduce the original dimensions. One popular feature extraction method is PCA. More details of the methods we used are discussed in Section ~\ref{sec2}.

In this paper, we propose a simple and intuitive wrapper method for dimension reduction in classification. The general idea is borrowed from sequential forward selection, which was proposed by \cite{Whitney1971}and is a commonly used technique for feature selection. Whitney proposed a heuristic method by using k-nearest neighbors (K-NN) and the leave-one-out method to sequentially examine the variables' performance. The techniques employ a suboptimum search procedure to determine the best subset of variables for classification. Instead of K-NN, we apply SVM as a single-variable classifier by considering each variable individually. Then, the performances is evaluated based on prediction rates. Finally, the selection criterion is adjusted to obtain the subset of variables with the best prediction results. This method provides a new criterion to reduce a complex dataset to lower dimensions. In addition, the geometric structural information embedded within high-dimensional datasets is extracted and maintained in the lower dimensions without distortion.

In most scientific endeavors, classification via supervised learning remains a fundamental tool for exploring knowledge via data analysis. However, classification typically requires tremendous effort and resources to label all study subjects and to collect many dimensions of their individual covariate information. As the labeling expense increases, it becomes unrealistic to label all study subjects. Thus, label information is often unavailable for large portions of the data, leaving only covariate information. Hence semi-supervised learning research is becoming increasingly popular and important \citep{seeger2000learning}. In our study, we focus on the performance of methods in a semi-supervised learning setting.

Using real microarray datasets, we compare the performance of stepwise SVM with the performance of widely used dimension reduction methods, including a filter method (correlation with SVM), embedded method (recursive feature elimination with random forest), and feature extraction method (PCA with SVM).

The rest of this article is organized as follows. In the Methods, we provide a brief overview of SVM and introduce the method we have developed. We present the comparison of the traditional methods and our approach on microarray datasets in the Results. A discussion and conclusions are offered in the last two sections.

\section{Method}
\label{sec2}
\subsection{Support Vector Machine (SVM)}
SVM \citep{Cortes1995, vapnik1998statistical} is a popular classification method in many fields. It is an effective supervised learning algorithm based on finding an optimal hyperplane that maximizes the margin of the training set. The hyperplane gives the largest minimum distance to separate the data points. By using a kernel, data points are mapped to a higher-dimensional space to improve the separation results.

Given a training set of n data nodes, $(x_i,y_i)$, $i=1, \ldots, n$, $y_i\in \{\pm 1\}$, we observe complete information in the sense that for every $x_i$ we know its class label, $y_i$. SVM is used as a classifier. The binary data nodes are separated by a hyperplane by
$f(x)=\sum_i \alpha_i y_i K(x_i,x)+b$, where $\alpha_i$ are positive constants and $b$ is the offset of the decision boundary from origin.

The Lagrange multiplier method and the conditions for optimality are applied to solve the problem in the dual space of Lagrange multipliers. The optimal $\alpha_i$ can be found by maximizing
$$\sum_{i=1}^n \alpha_i -\frac{1}{2}\sum_{i=1}^n\sum_{j=1}^n\alpha_i\alpha_jy_iy_j K(x_i,x_j)$$
subject to  $$\sum_{i=1}^n\alpha_i y_i,~\alpha_i\geq 0$$
If $y_i(\sum_i \alpha_i y_iK(x_i,x)+b)>1$, then all $i$ with $x_i$ are correctly classified.

The kernel functions project the nonlinear separable data into higher-dimensional space to perform linear separation. Commonly used kernel functions ($K(x_i,x)$) include the linear kernel: $x_i'x$; polynomial kernel: $(\gamma x_i'x + coef)^d$, where $d$ is the degree of the polynomial; Gaussian radial basis function (RBF): $exp(-\gamma \|x_i-x\|^2), \gamma \geq 0$; and sigmoid kernel: $tanh(\gamma x_i'x_j + coef)$.

\subsection{Stepwise SVM}
We propose a dimension reduction method called stepwise SVM. The main purpose of our method is to find the optimal subset of variables from a training dataset to achieve the best SVM prediction results. We start with an $n$ by $p$ dataset $X$ consisting of $p$ variables with $n$ data nodes. Each data node is labeled with $Y \in \{\pm 1\}$  (e.g., target disease or other). For data nodes that have more than two labels, the one-against-one technique is used by fitting all possible binary sub-classifiers and predicting the label with a voting mechanism. The goal is to find a subset of $q$ variables, which  $q<p$, by evaluating the performance of each variable and to use these q variables in an SVM classifier to obtain the maximum prediction performance.

The following is the algorithm for stepwise SVM:
\begin{itemize}
\item[1.] Perform simple random sampling without replacement to obtain a training set and a testing set.
\item[2.] Train an SVM with each variable to obtain $p$ SVM prediction models.
\item[3.] Calculate the apparent error rate (APR) for each SVM model and sort the rates.
\item[4.] Select an APR threshold. Remove variables from step 3 that have APRs greater than the threshold.
\item[5.] Train an SVM with the reduced dataset and calculate the accuracy.
\item[6.] Repeat steps 4 and 5 to adjust the threshold to find the subset with q variables with the highest accuracy.
\end{itemize}

\subsection{Other Methods}
\subsubsection{Filter: correlation}
The Pearson correlations, $r$, of paired variables are calculated and a threshold is set as a variable selection criterion. If a $|r|$ between two variables is larger than the threshold, then the variable with higher mean values is removed. The remaining variables are then trained with an SVM. In our study, the thresholds used to check for the best performance are 0.7, 0.75, 0.8, 0.85, 0.9, and 0.95.
\subsubsection{Embedded: Random Forest}
Random Forest was proposed by \cite{breiman2001}. The method randomly samples with replacement q variables to construct multiple decision trees for the training set. Each tree produces a classification result, and the label with the most votes is the prediction result. We used 500 decision trees and recursive feature elimination to select the best subset of variables.
\subsubsection{Feature extraction: PCA}
PCA was first formulated by Pearson \citep{Pearson} as a linear dimension reduction technique. The dimensions of data are reduced by linearly transforming the original variables into uncorrelated variables. The first several principal components explain most of the variance and represent the major structure of a datum. Principal components with lower variance represent noise and can be removed for subsequent analysis  \citep{kambhatla1997dimension}. Here, we fit SVMs with PCs. The first model is fitted by PC1, and the prediction results are obtained. The second model is fitted by PC1 and PC2 and the prediction results are obtained, and so on. Finally, the prediction results are compared to determine the best combination of PCs.

\section{Result}
\label{sec3}
\subsection{Data Collection}
\begin{figure}[H]
\includegraphics[width=\textwidth]{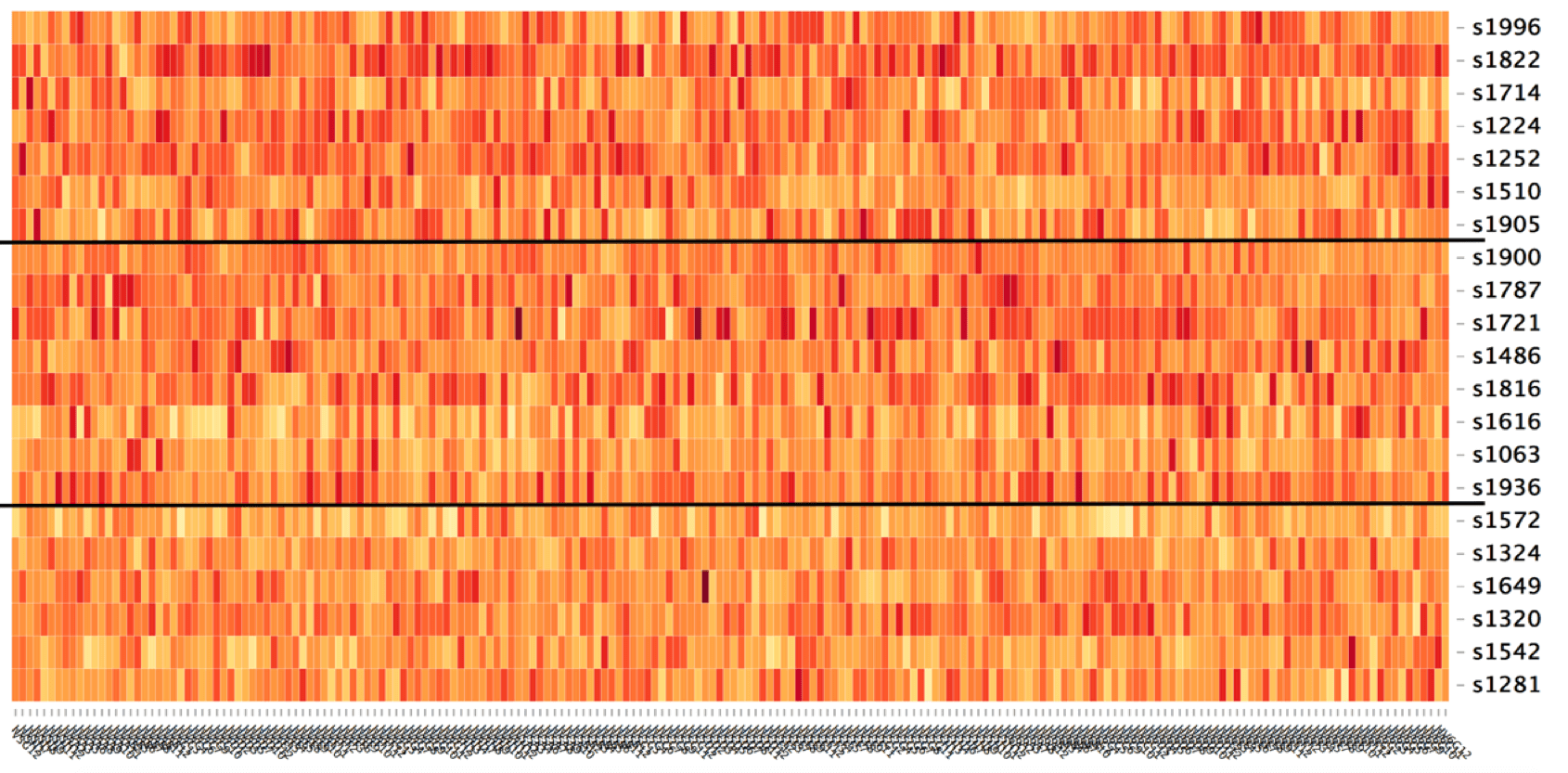}
\caption{ Heatmap of the relationships for Hedenfalk's partial dataset. The top section shows BRCA1 subjects, the middle section shows BRCA2 subjects, and bottom section shows others.}
\label{fig:1}
\end{figure}

We apply the method to five datasets. The first dataset is from a breast cancer microarray study performed by \cite{Hedenfalk2001}.  The data include information about breast cancer mutation in the BRCA1 and BRCA2 genes for 21 patients, 7 with BRCA1 mutations, 8 with BRCA2 mutations, and 6 with other types of mutations. The 3226 gene expression ratios are used as the variables. Figure ~\ref{fig:1} shows the relationship among cancer subtypes. Differences between BRCA1 and BRCA2 are difficult to find.

\begin{figure}[H]
  \centering
  \includegraphics[width=\textwidth]{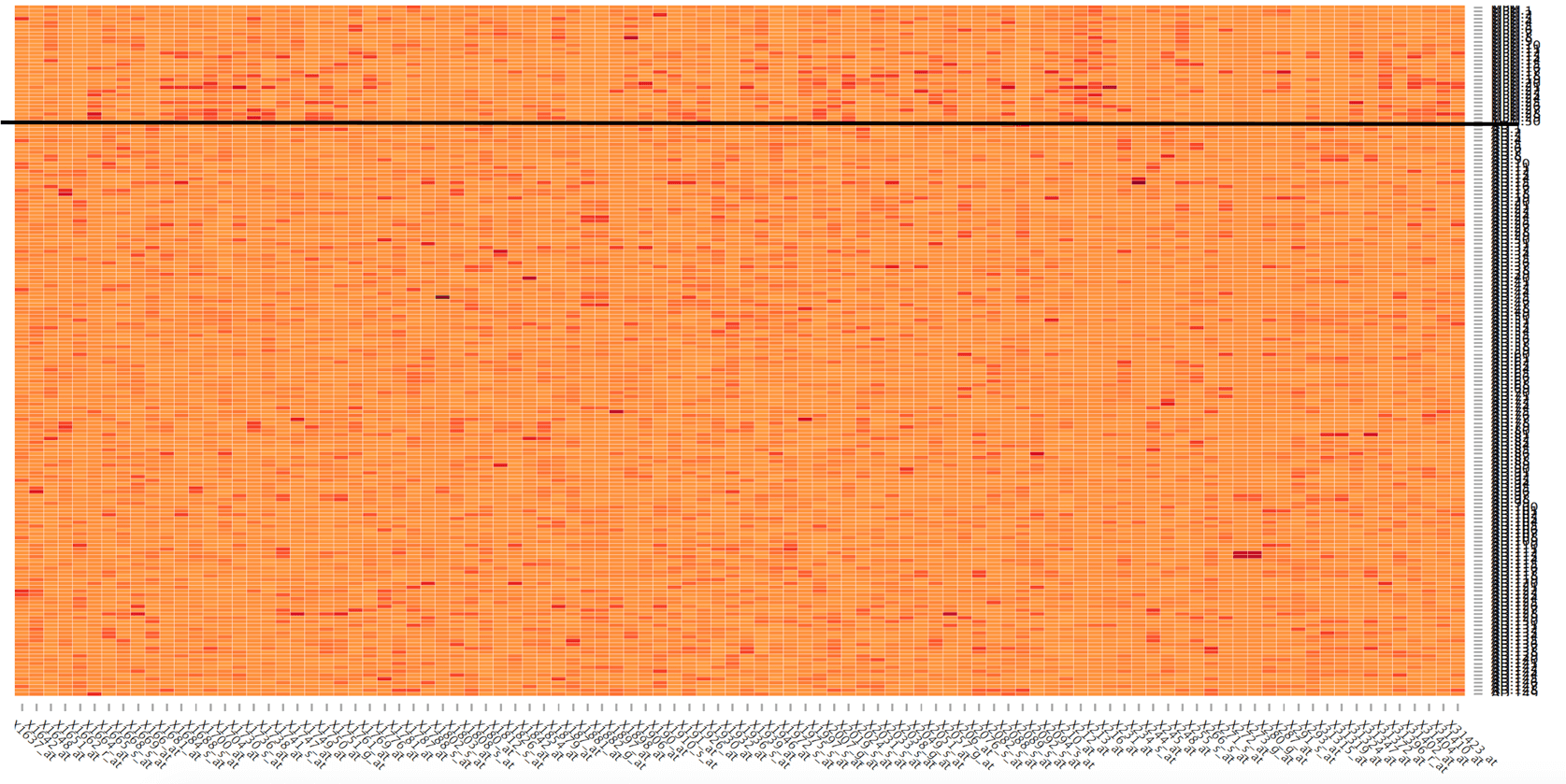}\\
  \caption{Heatmap of the relationships for Gordon's dataset. The top section shows MPM samples, and the bottom section shows ADCA samples.}
  \label{fig:2}
\end{figure}

The second gene expression dataset is from \cite{Gordon2002}. The data contain 31 malignant pleural mesothelioma (MPM) samples and 150 adenocarcinoma (ADCA) samples, with the expression of 1626 genes for each sample. Figure ~\ref{fig:2} shows the relationships for each cancer type. No pattern is observed in the non-reduced dataset.

\begin{figure}[H]
  \centering
  \includegraphics[width=\textwidth]{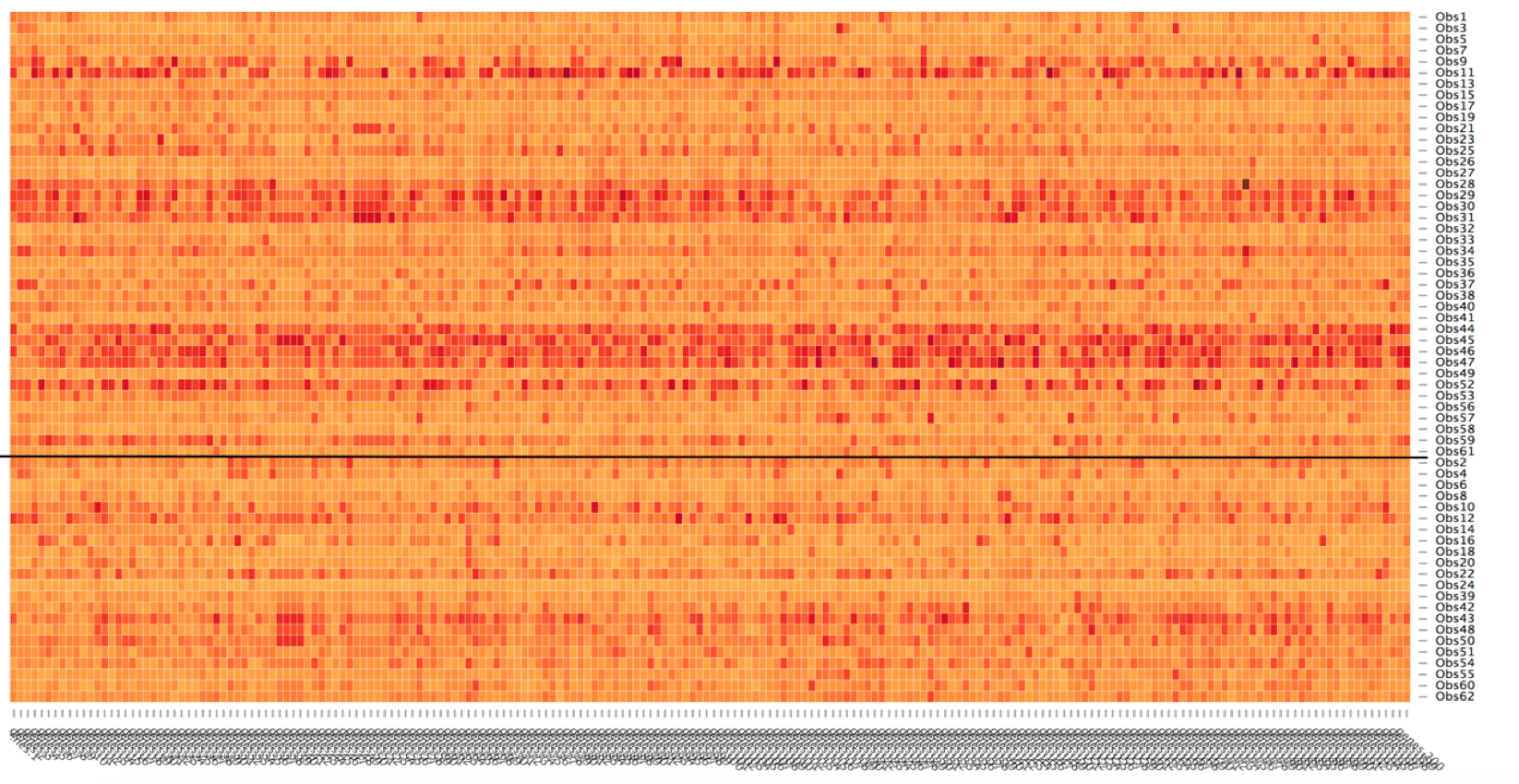}\\
  \caption{Heatmap of the relationships Alon's dataset. The top section shows tumor samples, and the bottom section shows normal samples.}
  \label{fig:3}
\end{figure}

The third set of microarray data is \cite{Alon1999}. The data include the expression of 2000 genes from 62 samples. These samples can be divided into two groups: 22 normal tissues and 40 tumor tissues. Figure ~\ref{fig:3} shows the relationships for type. The differences between tumor and normal samples are obvious.

\begin{figure}[H]
  \includegraphics[width=\textwidth]{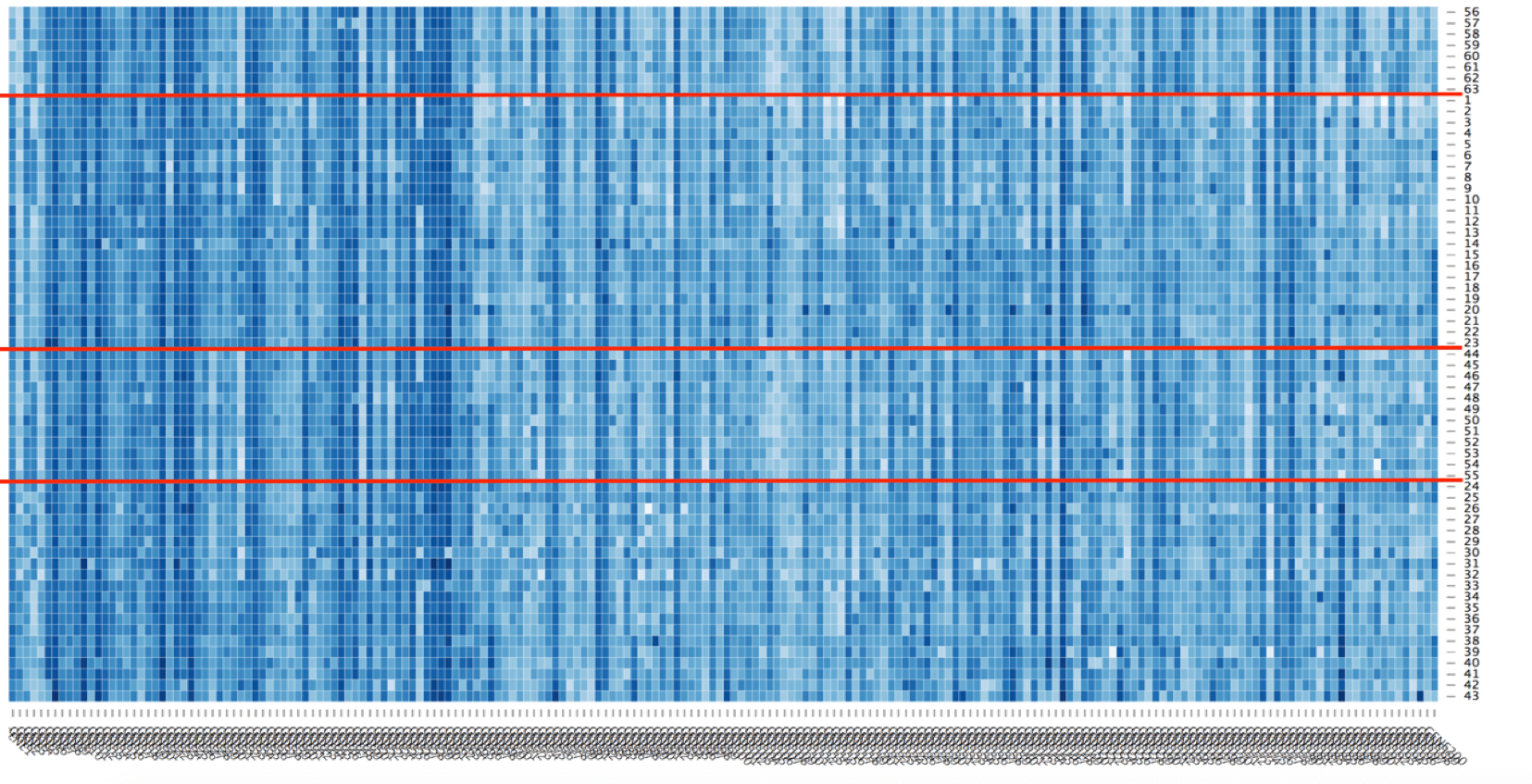}\\
  \caption{Heatmap of the relationships for Khan's partial dataset. The top to bottom sections are NHL, EWS, NB, and RMS samples.}
  \label{fig:4}
\end{figure}

The fourth dataset is from \cite{Khan2001},  and contains information about 2308 genes for 63 subjects with 4 different cancer types: 23 cases of Ewing sarcoma (EWS), 8 cases of non-Hodgkin lymphoma (NHL), 12 cases of neuroblastoma (NB), and 20 cases of rhabdomyosarcoma (RMS). The relationship of each cancer type is shown in Figure ~\ref{fig:4}.

\begin{figure}[H]
  \includegraphics[width=\textwidth]{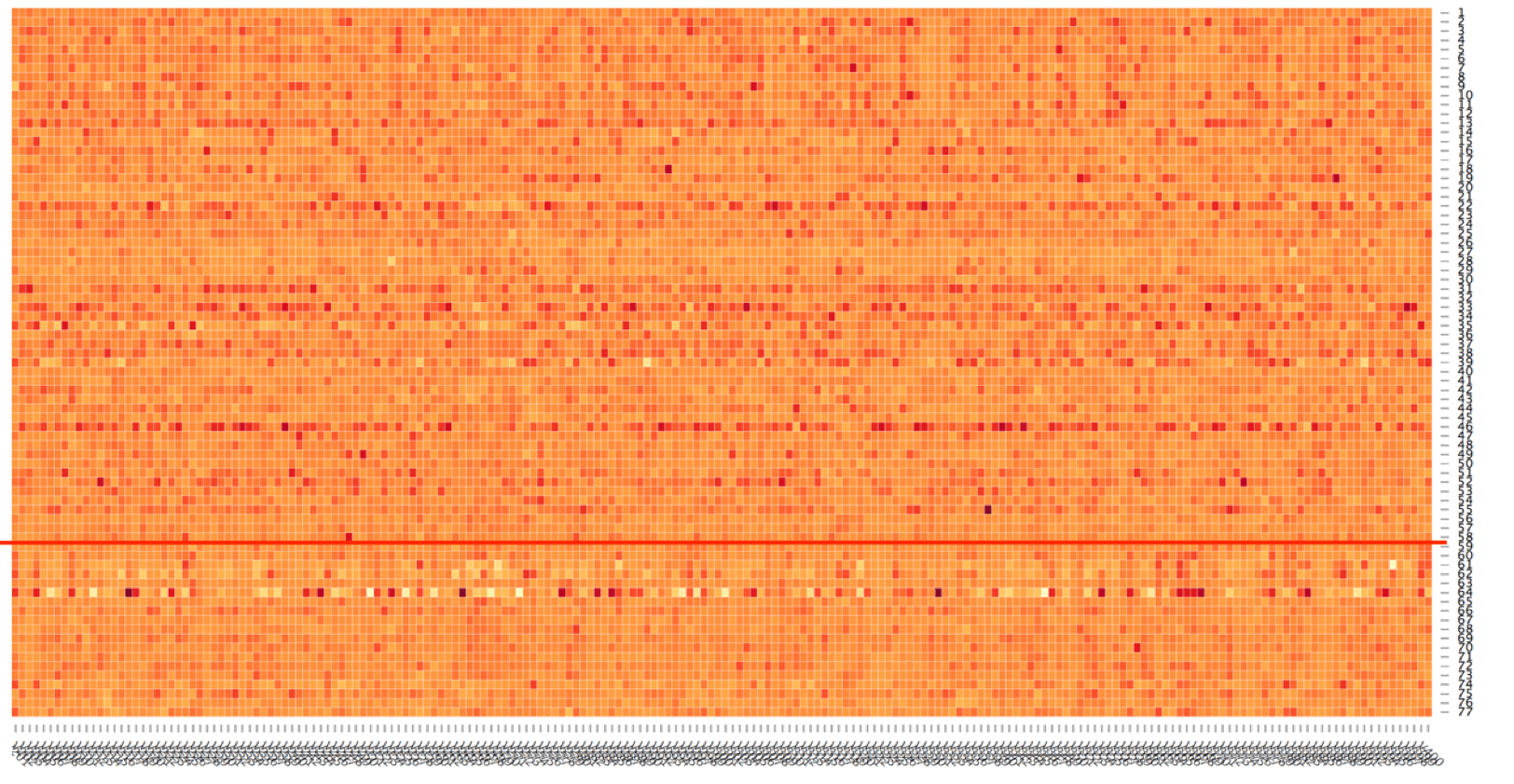}\\
  \caption{Heatmap of the relationships for Shipp's partial dataset. The top and bottom sections are DLBCL and FL samples, respectively.}
  \label{fig:5}
\end{figure}

The fifth dataset is from \cite{Shipp2002} and contains gene expression information for 7129 genes from 58 diffuse large B cell lymphoma (DLBCL) patient samples and 19 follicular lymphoma (FL) patient samples. The relationship of each tumor type is shown in Figure ~\ref{fig:5}. The pattern for each sample is not distinct.

\subsection{Supervised Learning Results}
\begin{figure}[H]
\includegraphics[width=\textwidth]{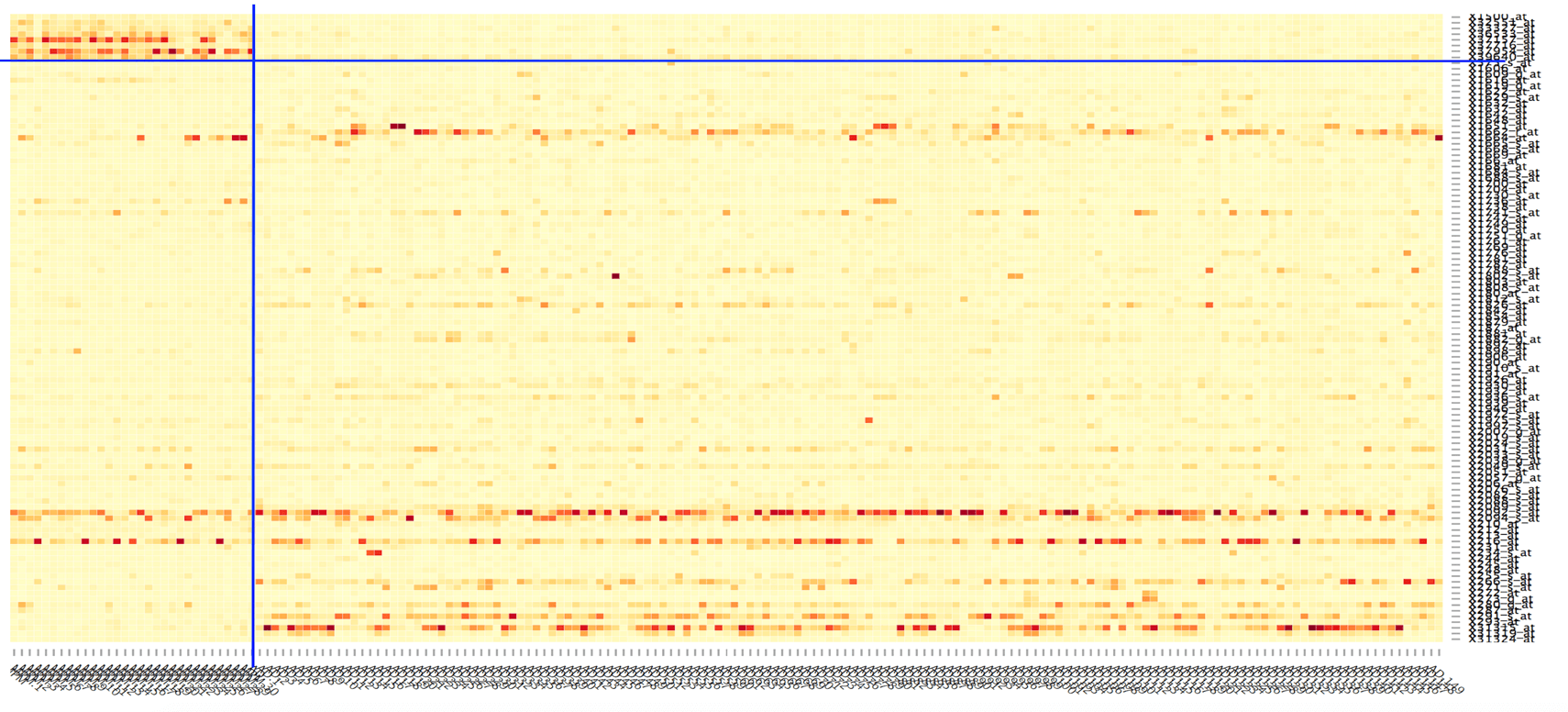}
\caption{Heatmap of the relationships between the subjects and selected and unselected variables for Gordon's dataset. The top section represents the selected variables.}
  \label{fig:6}
\end{figure}

\begin{figure}[H]
\includegraphics[width=\textwidth]{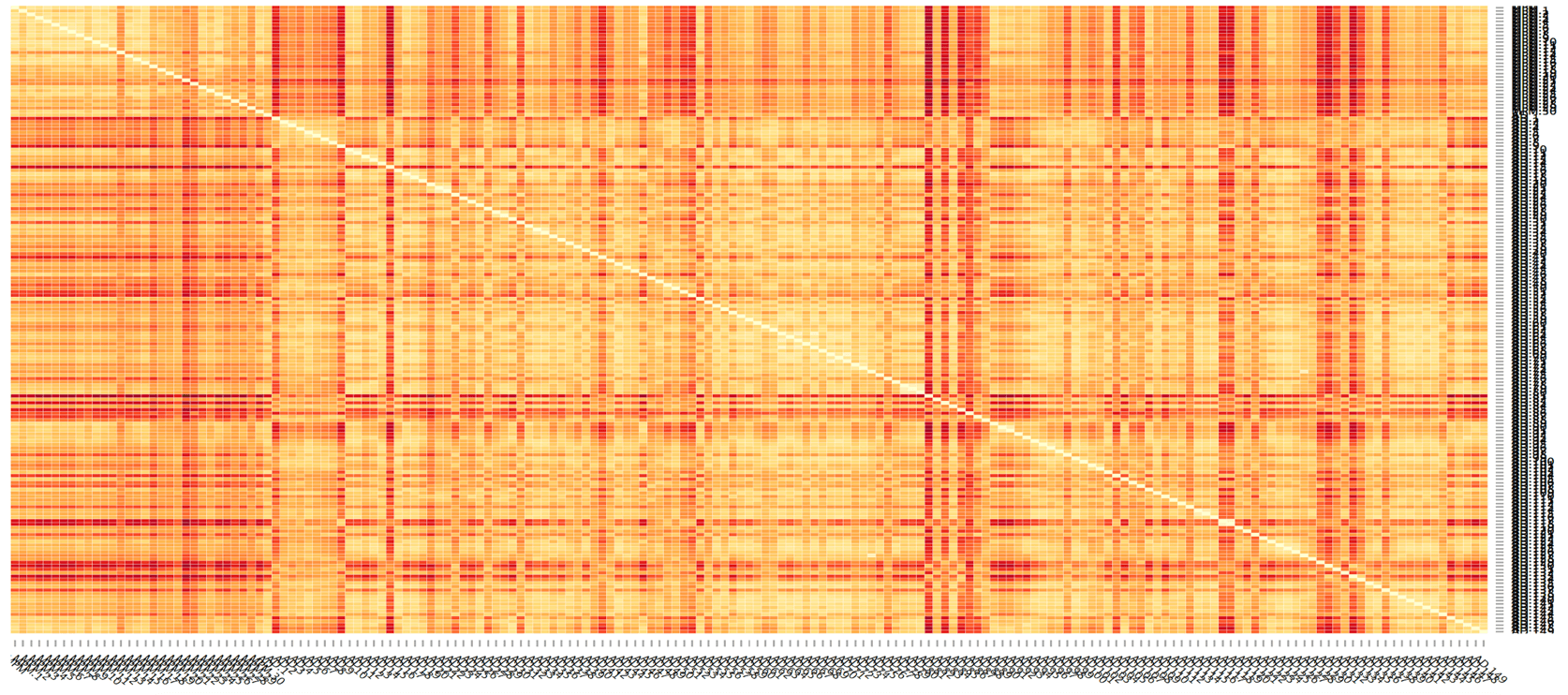}
\caption{Illustration of subject similarity in the original Gordon's dataset.}
  \label{fig:7}
\end{figure}

\begin{figure}[H]
\includegraphics[width=\textwidth]{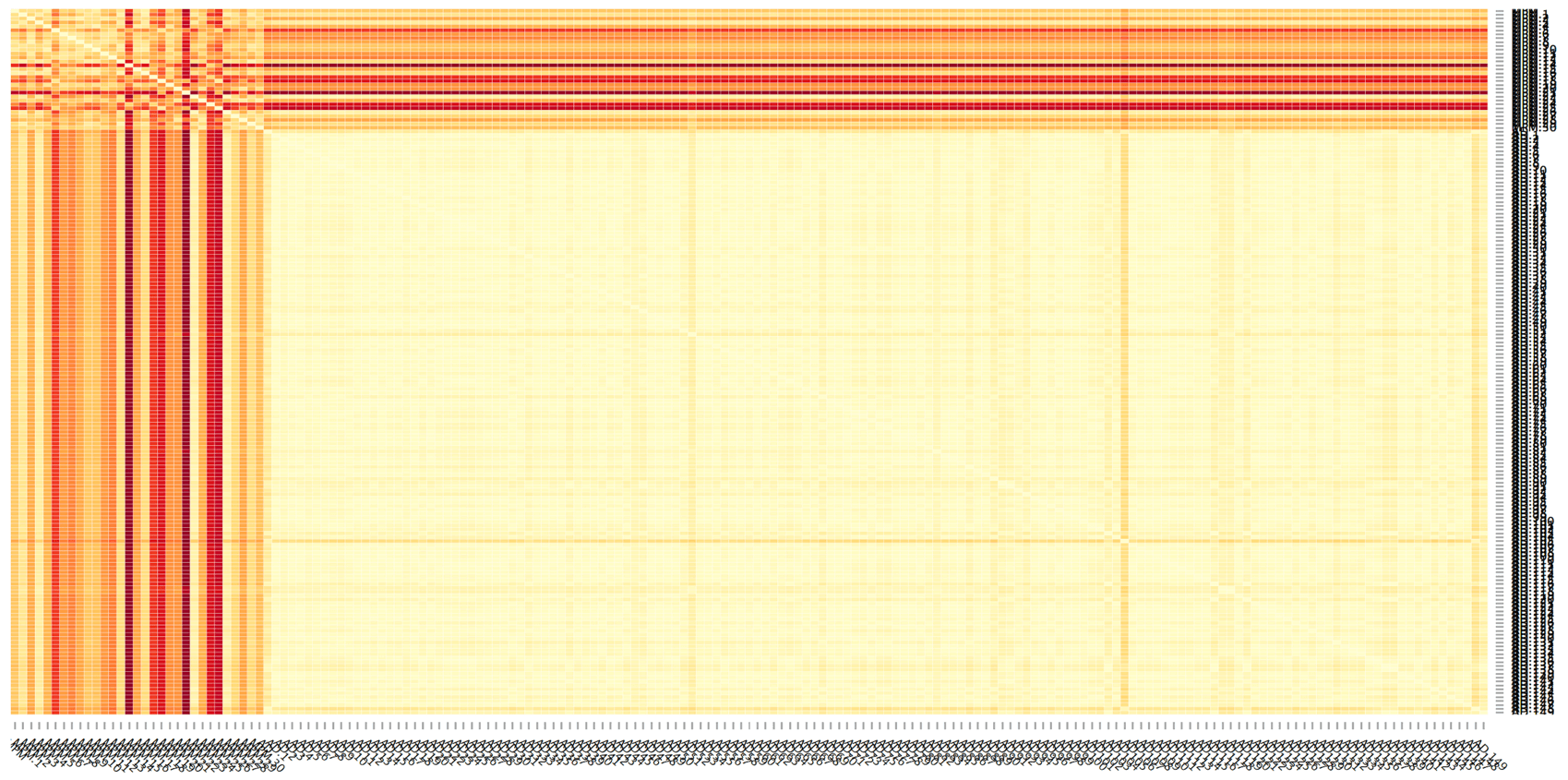}
\caption{Illustration of subject similarity in the reduced Gordon's dataset.}
  \label{fig:8}
\end{figure}

To verify that the proposed method is suitable for feature selection and classification, we apply the method to the data from Gordon et al. and Khan et al. In Gordon et al., the APR threshold is $\frac{6}{181}\approx 0.0331$, and 9 of 1626 variables are retained in the final dataset. These variables are selected as the best subset of variables for prediction. Figure ~\ref{fig:6} illustrates the relationship of subjects and selected variables and unselected variables. There is an obvious pattern for MPM samples: the selected variables are darker than the ADCA samples.
Figure ~\ref{fig:7} and ~\ref{fig:8} show the similarity of subjects in the original and reduced datasets. We use the Euclidean distance to calculate the dissimilarity of subjects. The darker the color is, the more dissimilar the subjects. The relationships between selected genes and different types of sample are clear in the reduced dataset. These selected genes can differentiate sample types much better than the whole genes can.

\begin{figure}[H]
\includegraphics[width=\textwidth]{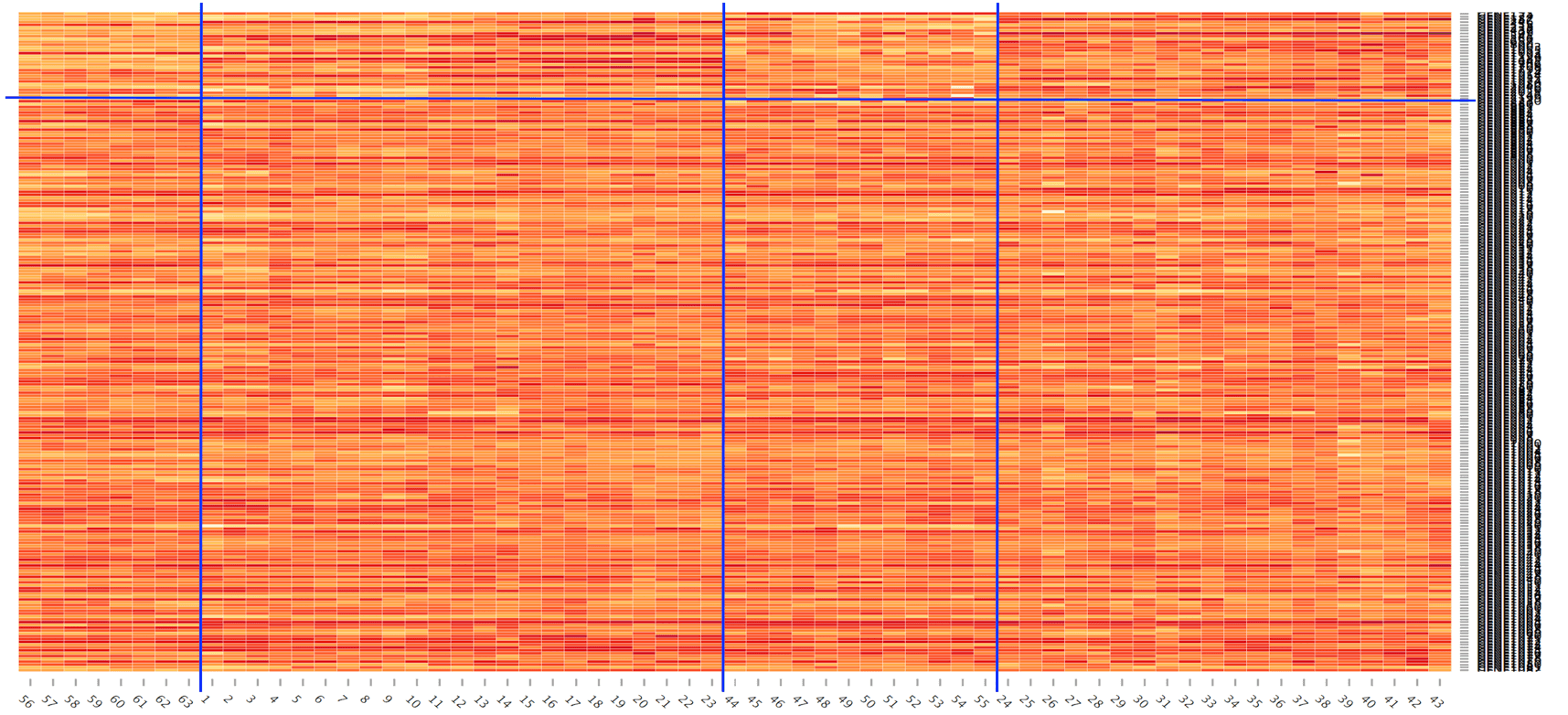}
\caption{Heatmap of the relationships between subjects and the selected and unselected variables for Khan's dataset. The top section shows the selected variables.}
  \label{fig:9}
\end{figure}

\begin{figure}[H]
\includegraphics[width=\textwidth]{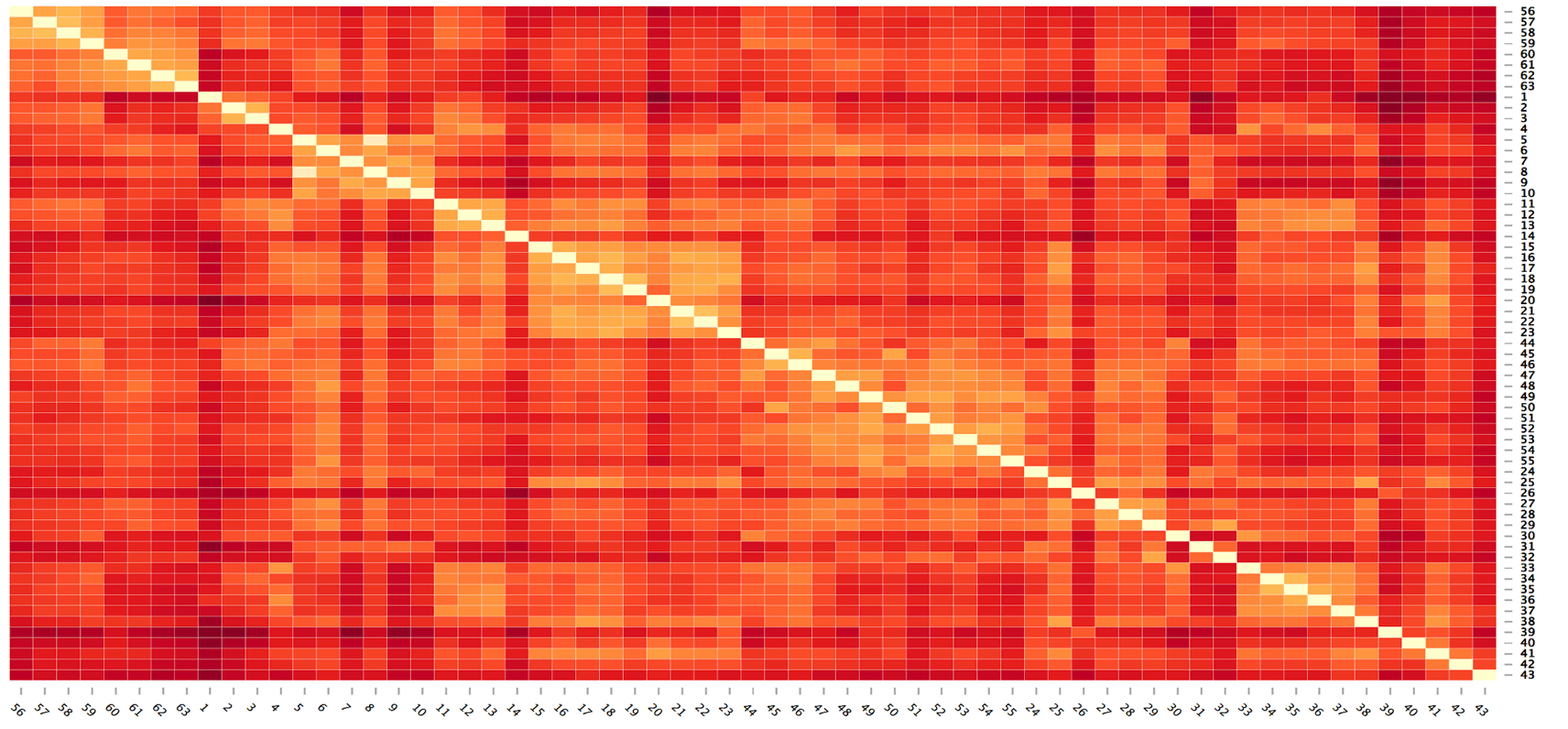}
\caption{Illustration of subject similarity in the original Khan's dataset.}
  \label{fig:10}
\end{figure}

\begin{figure}[H]
\includegraphics[width=\textwidth]{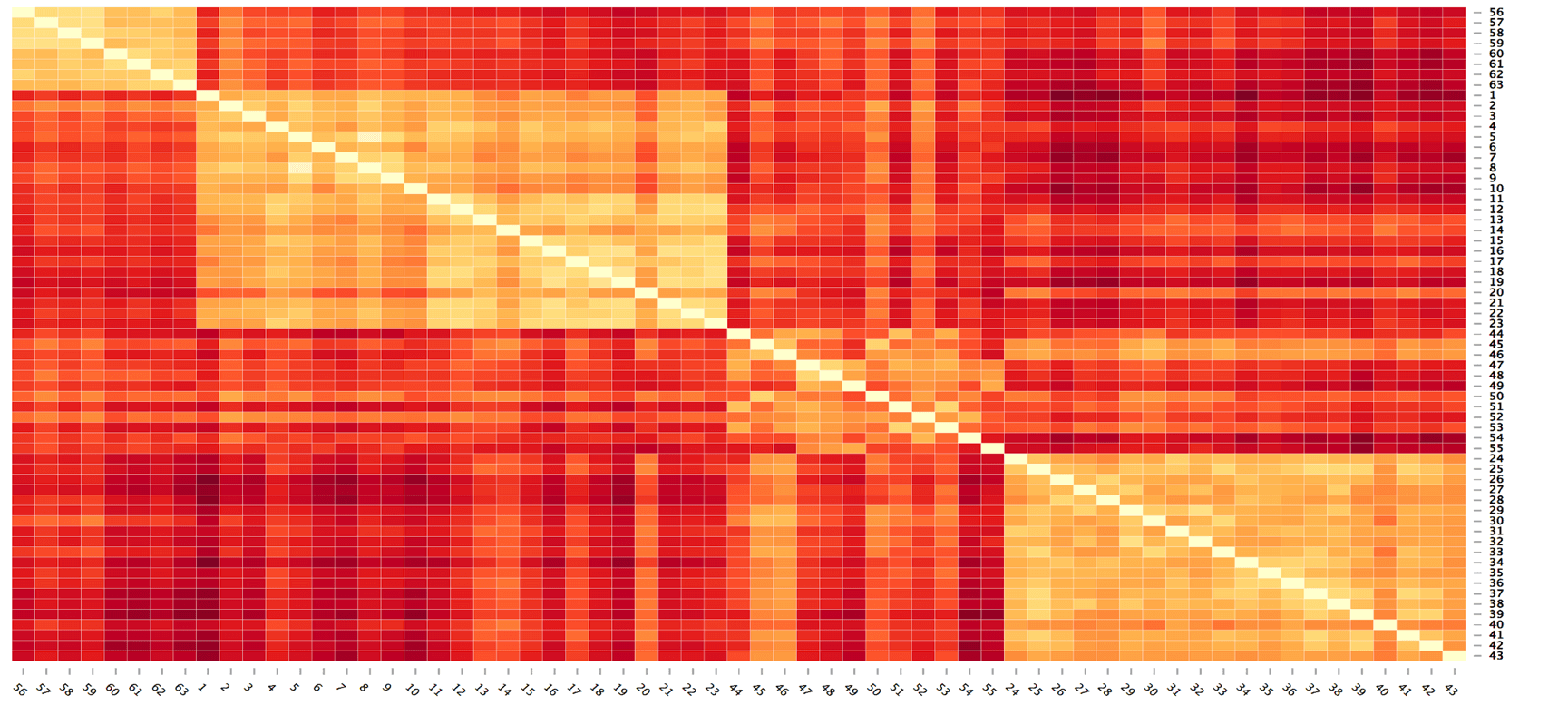}
\caption{Illustration of subject similarity in the reduced Khan's dataset.}
  \label{fig:11}
\end{figure}

In Khan et al., the APR threshold is $\frac{23}{63}\approx 0.3651$, and 32 of 2308 variables are selected for the final dataset. The relationships between subjects and the selected and unselected variables are shown in Figure ~\ref{fig:9}. The similarity of subjects in the original and reduced datasets is shown in Figure ~\ref{fig:10} and Figure~\ref{fig:11}. In the heatmap, the block patterns for each type of subject are clear with the selected genes.

The variables selected by stepwise SVM better segregate the subject groups than do the whole variables. The dissimilarity of subjects groups is obvious in the reduced datasets. The geometry of the datasets is also maintained in the reduced variables. Therefore, stepwise SVM method successfully selects the important variables to produce improved predictions in supervised learning.

\subsection{Semi-Supervised Learning Results}
The method is now extended to semi-supervised learning. We randomly select half of the subjects within the same group as a training set and use the other half as a testing set. The stepwise SVM is trained on the training set to obtain the best subset of variables. Classification is then performed on the testing set with only the selected variables. We repeat the random sampling and variable selection procedures 100 times to produce 100 prediction results. The average prediction rate is used as the final accuracy.

We apply our learning method to five datasets. A summary of the kernels used in the SVM and the APR thresholds used as the selection criterion can be found in Table ~\ref{table:1}.  Four different kernels are used in the experiments, and the one with the best prediction results is reported here.

We also compare the results with the dimension reduction methods that are commonly used in semi-supervised learning. The SVM is used as the classifier to generate the prediction accuracy. The comparison of the results is presented in Table ~\ref{table:2}, where the superscripts represent the ranking of the algorithms' performance.

First, we compare the results of the reduced data and the original whole data. The accuracy is clearly improved in the reduced datasets. Then, by comparing our method with the other dimension reduction methods, we find that for PCA, the prediction results are better on the Gordon et al. and Alon et al. datasets but are worse than those of the original dataset on the other datasets. Correlation as the dimension reduction method improves the results for the Hedenfalk et al. and Alon et al. datasets. RF-RFE does not perform well in our examples. Clearly, the performance of our method is more stable than that of the other methods.

\begin{table}[ht!]
\caption{Model description} 
\centering
\begin{tabular}[l]{l|ccccc}
                           & Hedenfalk & Gordon & Alon& Khan & Shipp    \\ \hline
Threshold                  & 3/11      & 4/91   & 6/31 & 12/32  & 10/39      \\
Kernel (variable selected) & \multicolumn{5}{c}{RBF}                              \\
Kernel (predict model)     & linear    & RBF    & RBF  & linear & linear \\[0.5ex]
\end{tabular}
\label{table:1} 
\end{table}

\begin{table}[ht!]
\caption{Classification accuracy for different examples (unit in \%)} 
\centering

\begin{tabular}{l|lllll}
          & Stepwise SVM                  & Original Data & PCA   & correlation & RF-RFE \\ \hline
Hedenfalk & $70.20^{(2)}$ & 68.60         & 68.10 & 70.50       & 67.90  \\
Gordon    & $98.61^{(2)}$ & 95.89         & 99.23 & 95.72       & 97.89  \\
Alon      & $80.58^{(3)}$ & 80.29         & 81.55 & 81.39       & 80.13  \\
Khan      & $98.65^{(1)}$ & 95.29         & 92.23 & 95.97       & 98.45  \\
Shipp     & $95.00^{(1)}$ & 94.97         & 91.76 & 95.00       & 91.32  \\
\end{tabular}
\label{table:2} 
\end{table}

\section{Discussion and Conclusion}
In this paper, we propose a dimension reduction method, stepwise SVM, based on individual variables' prediction quality before performing classification. Irrelevant variables are excluded to reduce the noise and to improve the performance of the classifier. Stepwise SVM is similar to stepwise regression. Through our learning rules, we have efficiently classified most of the datasets. The proposed method is effective compared with other dimension reduction techniques. The stable performance of stepwise SVM is shown in Section 3.

However, stepwise SVM has several limitations. First, the basic statistical classification in the framework of the SVM limits this approach to previously labeled data. When performing unsupervised learning, clustering methods may have to be used first to group the subject and to assign labels. Second, stepwise SVM considers variables one by one; therefore, variable dependencies and interactions are ignored when performing selection. Third, stepwise SVM produces suboptimal results. Fourth, substantial computation time is consumed to obtain the best subset of variables. The performance of computation time of the methods used in this study from best to worst is PCA, correlation, stepwise SVM, and RF-RFE.

Stepwise SVM is simple and intuitive, and the results are easily obtained by software. SVM is also widely used for classification in many fields. Thus, stepwise techniques can be easily applied and extensively used to reduce dimensions or to select important variables for prediction in numerous disciplines, such as medicine, psychology, and marketing.

\newpage





\newpage
\bibliographystyle{elsarticle-num-names}

\bibliography{stepsvm}
\biboptions{authoryear}

\end{document}